\ifx\pdfminorversion\undefined\else\ifnum\pdfminorversion<7\relax\pdfminorversion=7\relax\fi\fi

\documentclass[11pt]{article}
\usepackage{floatrow}
\usepackage{geometry}
\usepackage[affil-it, auth-sc]{authblk}
\usepackage[utf8]{inputenc}

\usepackage{caption}
\usepackage{subcaption}

\usepackage{booktabs}

\usepackage{graphicx}
\usepackage{amssymb}
\usepackage{amsthm}
\usepackage{amsmath}
\usepackage{color, soul}
\usepackage{verbatim}
\usepackage{lineno}
\usepackage{todonotes}
\presetkeys{todonotes}{inline, color=blue!30, size=\small}{}

\definecolor{darkred}{rgb}{0.9, 0.0, 0.0}
\definecolor{darkgreen}{rgb}{0.0, 0.5, 0.0}
\usepackage[pdftex,colorlinks,bookmarks,linkcolor=darkred, citecolor=darkgreen]{hyperref}

\usepackage[ampersand]{easylist}

\usepackage{bm}
\usepackage{feynmf}

\usepackage[sort&compress,numbers]{natbib}
\usepackage{doi}

\usepackage{slashed}

\usepackage{eso-pic}

\allowdisplaybreaks

\usepackage{graphicx}% Include figure files
\usepackage{dcolumn}% Align table columns on decimal point
\usepackage{bm}% bold math
\usepackage[english]{babel}
\usepackage[utf8]{inputenc}
\usepackage{float}
\usepackage{hyperref}
\usepackage{amssymb}   % for math
\usepackage{amsmath}   % for math
\usepackage{color}
\usepackage{comment}

\def\fAtbar{\bar{f}_{A3}}

\begin{document}

\AddToShipoutPictureFG*{
    \AtPageUpperLeft{\put(-60,-75){\makebox[\paperwidth][r]{FERMILAB-PUB-24-0068-PPD-T,~LA-UR-23-32361}}}  
    }

\title{\Large\bf Constraints on new physics with (anti)neutrino-nucleon scattering data}

\author[1]{Oleksandr Tomalak}
\author[2]{Minerba Betancourt}
\author[3]{Kaushik Borah}
\author[3,2]{Richard J.~Hill}
\author[2]{Thomas Junk}
\affil[1]{Theoretical Division, Los Alamos National Laboratory, Los Alamos, NM 87545, USA \vspace{1.2mm}}

\affil[2]{Fermilab, Batavia, IL 60510, USA \vspace{1.2mm}}

\affil[3]{Department of Physics and Astronomy, University of Kentucky, Lexington, KY 40506, USA \vspace{1.2mm}}

\date{\today}

\maketitle

\begin{abstract}
New physics contributions to the (anti)neutrino-nucleon elastic scattering process can be constrained by precision measurements, with controlled Standard Model uncertainties. In a large class of new physics models, interactions involving charged leptons of different flavor can be related, and the large muon flavor component of accelerator neutrino beams can mitigate the lepton mass suppression that occurs in other low-energy measurements. We employ the recent high-statistics measurement of the cross section for $\bar{\nu}_\mu p \to \mu^+ n$  scattering on the hydrogen atom by MINERvA to place new confidence intervals on tensor and scalar neutrino-nucleon interactions: $\mathfrak{Re} C_T = -1^{+14}_{-13} \times 10^{-4}$, $|\mathfrak{Im} C_T| \le 1.3 \times 10^{-3}$, and $|\mathfrak{Im} C_S| =  45^{+13}_{-19} \times 10^{-3}$. These results represent a reduction in uncertainty by a factor of $2.1$, $3.1$, and $1.2$, respectively, compared to existing constraints from precision beta decay.
\end{abstract}

\newpage

\section{Introduction}
\label{sec:introduction}

Charged-current elastic (anti)neutrino-nucleon scattering is the basic interaction process underlying many fundamental neutrino measurements. At low (1-10~MeV) energies, it serves as the main detection channel of reactor antineutrinos~\cite{Reines:1959nc,DayaBay:2012fng,RENO:2012mkc,DoubleChooz:2011ymz,JUNO:2015zny,KamLAND:2002uet,Hayes:2016qnu,STEREO:2018rfh,NEUTRINO-4:2018huq,PROSPECT:2018dtt} and solar neutrinos~\cite{Davis:1968cp,Cleveland:1998nv,SNO:2001kpb,SNO:2002tuh,SNO:2021xpa,Super-Kamiokande:1998kpq,Super-Kamiokande:2001ljr,Super-Kamiokande:2002weg,Super-Kamiokande:2005mbp,Super-Kamiokande:2001ljr,Borexino:2008gab,Borexino:2013zhu,BOREXINO:2014pcl}. At higher energies, this process is the underlying reaction of quasielastic (anti)neutrino-nucleus scattering~\cite{Formaggio:2013kya,Mosel:2016cwa,Alvarez-Ruso:2017oui}.

Precise knowledge of (anti)neutrino-nucleon amplitudes at GeV energies is required for the analysis of current~\cite{NOvA:2007rmc,T2K:2011qtm} and future~\cite{Alion:2016uaj,Abi:2020evt,Hyper-Kamiokande:2016dsw} neutrino oscillation experiments, which aim to determine neutrino oscillation parameters, discover charge-parity violation in the lepton sector, and conclusively establish the ordering of neutrino masses. Beyond their important role as input to neutrino-nucleus cross sections, these neutrino-{\it nucleon} amplitudes can also be used directly for fundamental physics analyses. Nucleon-level amplitudes are free from nuclear modeling uncertainties, and recent advances have improved our knowledge of Standard Model predictions for these quantities: electron-proton scattering~\cite{Bernauer:2010wm,Bernauer:2013tpr,Xiong:2019umf} and muonic atom spectroscopy~\cite{Pohl:2010zza,Antognini:2013txn} measurements have improved constraints on nucleon vector form factors; and lattice-QCD computations~\cite{Kronfeld:2019nfb,Meyer:2022mix,Alexandrou:2020okk,Park:2021ypf,Djukanovic:2022wru,Jang:2023zts} are reaching the accuracy of experimental constraints on the nucleon axial-vector form factor from  deuterium bubble-chamber~\cite{Mann:1973pr,Barish:1977qk,Miller:1982qi,Baker:1981su,Kitagaki:1983px,Meyer:2016oeg}, pion electroproduction~\cite{Amaldi:1972vf,Brauel:1973cw,DelGuerra:1975uiy,DelGuerra:1976uj,Esaulov:1978ed,Bhattacharya:2011ah}, and muon capture~\cite{MuCap:2012lei, Hill:2017wgb} data.

Motivated by these advances, we consider the application of (anti)neutrino-nucleon scattering data as a probe of new physics in neutrino interactions. In a large class of models, new physics contributions to (anti)neutrino-nucleon processes are suppressed by $m_\ell/M$, where $m_\ell$ and $M$ denote the charged lepton and nucleon masses, respectively. Thus even very precise measurements of nuclear beta decay rates, where $m_\ell = m_e$, can lead to only mild new physics constraints. Muon-based observables can mitigate the lepton mass suppression. Related examples include using energy levels in muonic hydrogen to measure the proton electromagnetic radius~\cite{Pohl:2010zza}, and using the muon capture rate in muonic hydrogen to measure the nucleon axial-vector radius~\cite{Hill:2017wgb}. Since accelerator neutrino beams are predominantly muon flavor, this lepton mass enhancement is naturally present for (anti)neutrino-nucleon scattering measurements.

The MINERvA Collaboration has presented~\cite{MINERvA:2023avz} the first high-statistics measurement of the $\bar{\nu}_\mu p \to \mu^+ n$ cross section, using a kinematic selection to separate scattering events on hydrogen from those on carbon, in a hydrocarbon target. In this paper, we consider the most general parameterization of the charged-current elastic antineutrino-proton amplitude and determine the impact of MINERvA data on possible new physics contributions.

\section{Theoretical framework} \label{sec:framework}

We restrict attention to left-handed neutrino fields. In the limit of massless charged lepton, the Standard Model matrix element for charged-current elastic antineutrino-proton scattering can be written as~\cite{longpaper}
\begin{align}
T^{m_\ell = 0}_{\bar{\nu}_\ell p \to \ell^+ n} = \sqrt{2}\mathrm{G}_\mathrm{F} V^*_{ud} \, \overline{\bar{\nu}}_\ell \gamma^\mu \mathrm{P}_\mathrm{L} \ell^+ 
\times \bar{n} \left[ \gamma_\mu \left( \bar{g}_M + \bar{f}_A \gamma_5 \right) - \left( \bar{f}_2 + 2 \fAtbar \gamma_5 \right) \frac{{K}_\mu}{M} \right] p, \label{eq:CCQE_amplitude}
\end{align}
with the CKM matrix element $V_{ud}$, the Fermi coupling constant $\mathrm{G}_\mathrm{F}$, the averaged nucleon momentum $K_\mu = \left( k_\mu + k^\prime_\mu \right)/2$, and the averaged nucleon mass $M$. For massive charged leptons when $m_\ell \neq 0$, four additional invariant amplitudes are required,
\begin{equation}
T_{\bar{\nu}_\ell p \to \ell^+ n} = T^{m_\ell = 0}_{\bar{\nu}_\ell p \to \ell^+ n} + \sqrt{2}\mathrm{G}_\mathrm{F} V^\star_{ud} \frac{m_\ell}{M} \left[ \frac{\bar{f}_{T}}{4} \overline{\bar{\nu}}_\ell \sigma^{\mu \nu} \mathrm{P}_\mathrm{R} \ell^+ \, \bar{n} \sigma_{\mu \nu}  p - \overline{\bar{\nu}}_\ell \mathrm{P}_\mathrm{R} \ell^+ \,
\bar{n} \left( \bar{f}_3 + \bar{f}_P \gamma_5 - \frac{\bar{f}_R}{4} \frac{\gamma^\mu P_\mu}{M} \gamma_5 \right) p \right], \, \label{eq:CCQE_amplitudem}
\end{equation}
with the averaged lepton momentum $P_\mu = \left( p_\mu + p^\prime_\mu\right)/2$. All invariant amplitudes are functions of two kinematical variables: the crossing-symmetric variable $\nu = E_\nu / M - \tau - r_\ell^2$, with the neutrino energy $E_\nu$, and the momentum transfer $Q^2 = - \left( p - p^\prime \right)^2 = - \left( k - k^\prime \right)^2$, where we define
\begin{equation}
    \tau = \frac{Q^2}{4M^2}, \qquad r_\ell = \frac{m_\ell}{2M} \,.
\end{equation}
Electric and magnetic amplitudes $\bar{g}_E$ and $\bar{g}_M$ are defined in terms of the amplitudes $\bar{f}_1$ and $\bar{f}_2$ as
\begin{equation}
 \bar{g}_E = \bar{f}_1 - \tau \bar{f}_2, \qquad \bar{g}_M = \bar{f}_1 + \bar{f}_2 \,.
\end{equation}

The eight amplitudes $\bar{f}_{1,2,3,A,A3,P,R,T}$ represent the most general Lorentz invariant structure. In the Standard Model at tree level (i.e., neglecting QED radiative corrections), the invariant amplitudes $\bar{f}_T$ and $\bar{f}_{R}$ vanish, and the remaining six amplitudes are real-valued functions of the squared momentum transfer $Q^2$, corresponding to the general matrix element of the left-handed quark current~\cite{LlewellynSmith:1971uhs,Fatima:2018tzs,Fatima:2018wsy,Fatima:2021ctt,SajjadAthar:2022pjt}. The ``second-class" amplitudes $\bar{f}_3$ and $\bar{f}_{A3}$ are proportional to the small isospin-violating quark mass difference, $\left(m_u - m_d \right)/M$. QED radiative corrections contribute to all eight invariant amplitudes~\cite{Tomalak:2021hec,Tomalak:2022xup}. In the Standard Model, the invariant amplitudes $\bar{f}_3,~\bar{f}_P,~\bar{f}_R,$ and $\bar{f}_T$ enter physical observables with a factor of the charged lepton mass $m_\ell$. This property will hold for any new physics contribution described by a second-class quark current~\cite{LlewellynSmith:1971uhs,Wilkinson:2000gx,Day:2012gb,Cirigliano:2013xha}, and also holds for scalar and tensor interactions in the class of new physics models with Minimal Flavor Violation in the lepton sector~\cite{Cirigliano:2005ck}.\footnote{Exact proportionality to $m_\ell$ holds for the allowed quark-level scalar and tensor operators in the quasi-degenerate neutrino mass limit, $\Delta m^2 \ll m_\nu^2$, where $\Delta m^2$ denotes the difference of squared neutrino masses, and $m_\nu^2$ is the absolute neutrino mass scale, cf. Eqs.~(18) and (21) in Ref.~\cite{Cirigliano:2005ck}. Away from the quasi-degenerate limit, computable corrections involving neutrino mass and mixing parameters relate electron and muon flavor interactions.} We focus our attention on this case and compare constraints from the scattering data to the corresponding bounds from beta decay observables, which describe interactions involving the electron flavor. We note however that our results can be immediately translated to bounds on muon-specific neutrino interactions, which have no corresponding beta decay constraint.

Using the representation of Eqs.~(\ref{eq:CCQE_amplitude})~and~(\ref{eq:CCQE_amplitudem}), the charged-current elastic antineutrino-proton unpolarized scattering cross section in the laboratory frame is expressed as~\cite{LlewellynSmith:1971uhs,longpaper}
\begin{equation}
\frac{d\sigma}{dQ^2} \left( E_\nu, Q^2 \right) = \frac{\mathrm{G}_\mathrm{F}^2 |V_{ud}|^2}{2\pi} \frac{M^2}{E_\nu^2} \left[ \left( \tau + r_\ell^2 \right)A \left(\nu,~Q^2 \right) + \frac{\nu}{M^2} B \left(\nu,~Q^2 \right) + \frac{\nu^2}{M^4} \frac{C \left(\nu,~Q^2 \right)}{1+ \tau} \right] \,.\label{eq:xsection_CCQE}
\end{equation}
The structure-dependent quantities $A$, $B$, and $C$ are functions of $E_\nu$ and $Q^2$ and are expressed in terms of the invariant amplitudes as
\begin{align}
A &= \tau | \bar{g}_M |^2 - | \bar{g}_E |^2 + \left( 1+ \tau \right) | \bar{f}_A |^2 - r_\ell^2 \left( | \bar{g}_M |^2 + | \bar{f}_A + 2 \bar{f}_P |^2 - 4 \left( 1 + \tau \right) \left( |\bar{f}_P|^2 + |\bar{f}_3|^2\right)\right) \nonumber \\
& - 4 \tau \left( 1+ \tau \right) |\fAtbar|^2 + \frac{r_\ell^2}{4} \left( \nu^2 + 1 + \tau - \left(1 + \tau + r_\ell^2 \right)^2 \right) |\bar{f}_R|^2  - r_\ell^2 \left( 1 + 2 r_\ell^2 \right) |\bar{f}_T|^2 \nonumber \\
&- 2 r_\ell^2 \mathfrak{Re} \Big[ \left(  \bar{g}_E + 2 \bar{g}_M  - 2 \left( 1 + \tau \right) \fAtbar \right) \bar{f}_T^\star \Big] + r_\ell^2 \left( 1 + \tau + r_\ell^2 \right)  \mathfrak{Re}\Big [ \bar{f}_A \bar{f}_R^\star \Big] + 2 r_\ell^4 \mathfrak{Re}\Big [ \bar{f}_P \bar{f}_R^\star \Big]\,,\nonumber \\
B &= \mathfrak{Re} \Big[ 4 \tau \bar{f}^*_A \bar{g}_M - 4 r_\ell^2 \left( \bar{f}_A - 2 \tau \bar{f}_P \right)^*  \fAtbar - 4 r_\ell^2 \bar{g}_E \bar{f}_3^\star - 2 r_\ell^2 \left( 3 \bar{f}_A - 2 \tau \left( \bar{f}_P + \bar{f}_3\right) \right) \bar{f}_T^\star - r_\ell^4 \left( \bar{f}_T  + 2 \fAtbar  \right) \bar{f}_R^\star \Big]  \,,  \nonumber \\
C &= \tau |\bar{g}_M|^2 + |\bar{g}_E|^2 + \left( 1 + \tau \right) |\bar{f}_A|^2 + 4\tau \left( 1 + \tau \right) |\fAtbar|^2 + 2 r_\ell^2 \left( 1 + \tau \right) |\bar{f}_T|^2 - r_\ell^2 \left( 1 + \tau \right)  \mathfrak{Re} \Big[ \bar{f}_A \bar{f}_R^\star \Big] \,.
\end{align}

\section{Analysis}

Let us consider experimental constraints on the amplitudes $\bar{f}_{3,A3,R,T}$, which vanish up to small corrections in the Standard Model, as discussed above. We consider eight fits where we include the real or imaginary part of one additional amplitude, represented as
\begin{align} \label{eq:fp5_constraint}
   \bar{f}^j_i  \left( \nu, Q^2 \right)  &=  \frac{\mathfrak{Re} \bar{f}^j_i \left( 0 \right) + i \mathfrak{Im} \bar{f}^j_i \left( 0 \right)}{\left( 1 + \frac{Q^2}{\Lambda^2} \right)^2} \,.
\end{align}
For simplicity, we describe the $Q^2$ dependence of the new amplitudes using the dipole ansatz~\cite{Baker:1981su,Holstein:1984ga,AHRENS1988284,Day:2012gb} with the dipole parameter $\Lambda \approx 1~\mathrm{GeV}$. In each fit, only the real part or the imaginary part of one invariant amplitude is allowed to float. Standard Model amplitudes are allowed to float within their uncertainties.

For $\bar{f}_1$ and $\bar{f}_2$ at tree level, we take the nucleon vector form factors $F_1 \left( Q^2 \right)$ and $F_2 \left( Q^2 \right)$ from the fit of Ref.~\cite{Borah:2020gte} that included electron-proton and electron-deuteron scattering data, and constraints from the muonic hydrogen Lamb shift and neutron scattering length. For $\bar{f}_A$ and $\bar{f}_P$ at tree level, we take the nucleon axial-vector form factor $F_A(Q^2)$ from the fit of Ref.~\cite{Meyer:2016oeg} to deuterium bubble-chamber data, and the standard (partially conserved axial-vector current and pion pole dominance) ansatz for the pseudoscalar form factor:
\begin{equation} \label{eq:pseudoscalarFF}
 F_P \left( Q^2 \right) = \frac{2 M^2 F_A \left( Q^2 \right)}{m_\pi^2 + Q^2} \,. 
\end{equation}
Standard Model uncertainties from $\bar{f}_{1,2,A,P}$ are subleading compared to experimental errors in the current analysis. QED radiative corrections~\cite{Tomalak:2021hec,Tomalak:2022xup} can be incorporated~\cite{longpaper} but are also a subleading effect in the current analysis.

For the experimental data, we take the fifteen data points and the corresponding covariance matrix from the antineutrino-hydrogen cross-section measurement by the MINERvA Collaboration ~\cite{MINERvA:2023avz}. We integrate Eq.~(\ref{eq:xsection_CCQE}) over the incoming antineutrino NUMI flux, and put kinematical cuts on the recoil muon scattering angle $\theta_\mu \le 20^0$ and momentum $1.5~\mathrm{GeV} \le p_\mu \le 20~\mathrm{GeV}$~\cite{MINERvA:2023avz}. Floating the normalization of one of the invariant amplitudes, we calculate fifteen cross-section predictions and the corresponding covariance matrix. Subsequently, for each invariant amplitude in Eq.~(\ref{eq:fp5_constraint}) we perform a log-likelihood minimization accounting for bin-to-bin correlations both in data and in theory and present $1\sigma$ ($68~\%$ confidence level) intervals for each parameter. We present plots with log-likelihood functions in the Supplementary Material.
\begin{center}
\begin{table}[t]
 \centering 
\renewcommand{\arraystretch}{1.3}
\begin{tabular}{|c|c|c|c|c|}
  \hline
& \multicolumn{1}{c|}{$\mathfrak{Re} \bar{f}_3$} & \multicolumn{1}{c|}{$\mathfrak{Re} \bar{f}_{T}$} &  \multicolumn{1}{c|}{$\mathfrak{Re} \fAtbar$} & \multicolumn{1}{c|}{$\mathfrak{Re} \bar{f}_R$}  \\  \hline
$\bar{\nu}p$ scattering & $88.4^{+33.5}_{-58.0}$ & $-0.5^{+5.0}_{-4.8}$ &  $-1.0^{+0.4}_{-0.3}$ \& $1.0^{+0.3}_{-0.4}$ & $-80.1^{+40.6}_{-26.0}$ \\ \hline
 beta decay & $0.0 \pm 1.8$~\cite{Hardy:2020qwl} & $-9.3\pm 10.3$~\cite{Gonzalez-Alonso:2018omy} & $0.0 \pm 0.075$~\cite{Day:2012gb}  & \\ \hline
 \end{tabular}
 \caption{$68\%$ shortest Bayesian credibility intervals with uniform prior for the real parts of the amplitudes from the $\bar{\nu} p$ fit are compared to constraints from beta decay. \label{tab:results_re}}
\renewcommand{\arraystretch}{1}
\end{table}
\end{center}

\begin{center}
\begin{table}[t]
\centering
\renewcommand{\arraystretch}{1.3}
\begin{tabular}{|c|c|c|c|c|c|}
  \hline 
&
\multicolumn{1}{c|}{$\mathfrak{Im} \bar{f}_3$} & 
\multicolumn{1}{c|}{$\mathfrak{Im} \bar{f}_{T}$} &  
\multicolumn{1}{c|}{$| \mathfrak{Im} \fAtbar |$} & 
\multicolumn{1}{c|}{$| \mathfrak{Im} \bar{f}_R |$} \\ \hline 
  $\bar{\nu}p$ scattering & $-82.1_{-23.8}^{+34.6}$ \& $82.1^{+23.8}_{-34.6}$ & $0.0\pm4.9$ & $1.00^{+0.29}_{-0.43}$ & $69.9^{+20.9}_{-30.9}$  \\ \hline
  beta decay & $13.0 \pm 54.0$~\cite{Gonzalez-Alonso:2018omy} & $-1.9 \pm 15.4$~\cite{Gonzalez-Alonso:2018omy} & & \\ \hline
 \end{tabular}
 \caption{$68\%$ shortest Bayesian credibility intervals with uniform prior for the imaginary parts of the amplitudes from the $\bar{\nu} p$ fit are compared to constraints from beta decay.\label{tab:results_im}}
\renewcommand{\arraystretch}{1}
\end{table}
\end{center}

\section{Results}

In Tables~\ref{tab:results_re} and \ref{tab:results_im}, we provide results of our fits for amplitudes $\bar{f}_{3,A3,R,T}$ at $Q^2=0$, and compare to the bounds obtained from beta decay measurements~\cite{Hardy:2004id,Severijns:2006dr,Hardy:2008gy,Kozela:2011mc,Day:2012gb,Hardy:2014qxa,Gonzalez-Alonso:2018omy,Hardy:2020qwl}. As the tables illustrate, the MINERvA antineutrino-proton data provide improved bounds on $\mathfrak{Re} \bar{f}_T$, $\mathfrak{Im} \bar{f}_T$, and $\mathfrak{Im} \bar{f}_3$ in the assumption of lepton flavor dependence from Section~\ref{sec:framework}. Identifying Lee-Yang coefficients~\cite{Lee:1956qn,Gonzalez-Alonso:2018omy} as $C_T = \left( m_e/2M \right) \bar{f}_T\left(0 \right)$ and $C_S = \left( -m_e/M \right) \bar{f}_{3}\left(0 \right)$, these new bounds translate to $\mathfrak{Re} C_T = -1^{+14}_{-13} \times 10^{-4}$, $|\mathfrak{Im} C_T| \le 1.3 \times 10^{-3}$, and $|\mathfrak{Im} C_S| =  45^{+13}_{-19} \times 10^{-3}$ compared to the fit of beta decay observables $\mathfrak{Re} C_T = -0.0025\left( 28 \right)$, $\mathfrak{Im} C_T = -0.0005\left( 42 \right)$, and $\mathfrak{Im} C_S = -0.007\left( 30 \right)$~\cite{Gonzalez-Alonso:2018omy}. Based on the length of the $1\sigma$ confidence interval, the new bounds represent an improvement in sensitivity to $\mathfrak{Re} C_T$, $|\mathfrak{Im} C_T|$, and $|\mathfrak{Im} C_S|$ by factors $2.1$, $3.2$, and $1.2$, respectively.

Using the scalar charge $g_S = 1.02 \pm 0.10$ and the tensor charge $g_T = 0.989 \pm 0.034$ at the renormalization scale $\mu = 2~\mathrm{GeV}$~\cite{Gonzalez-Alonso:2013ura,BMW:2014pzb,Gonzalez-Alonso:2018omy,Gupta:2018qil,FlavourLatticeAveragingGroupFLAG:2021npn}, the above results can alternatively be interpreted as direct constraints on muon-specific interactions: $\mathfrak{Re} \left( \varepsilon_S^{u d} \right)_{\mu \mu} = - 9.8_{-3.7}^{+6.4},$ $|\mathfrak{Im} \left( \varepsilon_S^{u d} \right)_{\mu \mu}| =  9.1^{+2.6}_{-3.8},$ $\mathfrak{Re} \left( \varepsilon_T^{u d} \right)_{\mu \mu} = 7^{+68}_{-71} \times 10^{-3},$ and $|\mathfrak{Im} \left( \varepsilon_T^{u d} \right)_{\mu \mu}| \le  7\times 10^{-2}$, in the notation of Refs.~\cite{Lee:1956qn,Jackson:1957zz,Severijns:2006dr,Cirigliano:2009wk,Bhattacharya:2011qm,Cirigliano:2012ab,Naviliat-Cuncic:2013ylu,Gonzalez-Alonso:2016etj,Falkowski:2017pss,Gonzalez-Alonso:2018omy,Falkowski:2019xoe,Falkowski:2020pma,Falkowski:2021bkq,Cirigliano:2023nol,Dawid:2024wmp}.

Compared to the existent constraint from the neutrino-deuterium scattering data of the experiment at the Brookhaven National Laboratory~\cite{Baker:1981su}, $|\mathfrak{Re} \fAtbar| \le 2.0~\left( 90\%~\mathrm{C.L.} \right)$, MINERvA provides a constraint of the same order of magnitude: $|\mathfrak{Re} \fAtbar| = 0.2-1.4~\left( 90\%~\mathrm{C.L.} \right)$. Sensitivity to this amplitude was also investigated in the bubble-chamber experiment at the Argonne National Laboratory~\cite{Barish:1977qk}.

\section{Discussion}

In this paper, we constrain low-energy nucleon contact interactions with neutrino scattering data, using the recent antineutrino-hydrogen measurement by the MINERvA Collaboration. We significantly improve the neutron beta decay constraint for the real and imaginary parts of the tensor interaction, and slightly improve the constraint for the imaginary part of the scalar interaction. We provide the first-ever constraints on the amplitudes $\mathfrak{Im}\fAtbar,~\mathfrak{Re}\bar{f}_R,$~and $\mathfrak{Im}\bar{f}_R$. Several directions related to this work are interesting to pursue:
\begin{itemize}
    \item 
    We have used the recent MINERvA data, which provides a complete error matrix. It is interesting to consider other possible datasets. Deuterium bubble-chamber measurements at ANL~\cite{Mann:1973pr,Barish:1977qk,Miller:1982qi} and BNL~\cite{Baker:1981su} were performed at lower neutrino energies where sensitivity to the scalar interactions increases. It is interesting to revisit these fits. The future DUNE/LBNF experiment may provide new high-statistics data for antineutrino-hydrogen scattering. This possibility has been investigated in the context of Standard Model axial-vector form factor determination~\cite{Petti:2023abz} and it is interesting to consider sensitivity to potential new physics. Neutrino-nucleus scattering may also be considered, bringing considerations of nuclear modeling~\cite{Belikov:1983kg}.  Enhanced sensitivity to pseudoscalar and tensor interactions was recently found for neutrino-nucleus cross sections on the oxygen in Ref.~\cite{Kopp:2024yvh}.

    \item 
    It is interesting to explore the implications of Minimal Flavor Violation beyond the quasi-degenerate neutrino mass limit, where exact proportionality to the lepton mass holds in Eq.~(\ref{eq:CCQE_amplitudem}), including corrections from electron-flavor neutrino flux on muon charged-current production.

    \item
    Uncertainties in our fits are dominated by MINERvA experimental errors. We have taken Standard Model vector and axial-vector form factors from fits to experimental data assuming the absence of new physics~\cite{Borah:2020gte,Meyer:2016oeg}. It is interesting to consider external theoretical input, e.g., from lattice QCD, for the Standard Model amplitudes, in order to consistently separate potential new physics. Currently, this would result in a loss of sensitivity from lattice uncertainties in the vector form factors.   With more precise data, it is also interesting to include a more complete form factor description beyond the dipole ansatz employed in Eq.~(\ref{eq:fp5_constraint}).

    \item   
    For amplitudes $\bar{f}_{3,A3,R,T}$, we have performed separate fits taking a single nonvanishing amplitude (and either its real or imaginary part).  It may be interesting to investigate simultaneous fits of multiple amplitudes in the context of specific new physics models. It is also interesting to investigate the range of models for which comparison can be made between neutrino scattering at $\sim$GeV energies and LHC constraints at much higher energy, which assume the applicability of quark-level contact interactions to high momentum scales. Tensor interactions with the muon flavor can be constrained at FASER$\nu$~\cite{Falkowski:2021bkq} but with a looser bound than we have found. Our bounds on $\left( \epsilon_{S,T}^{ud} \right)_{\mu\mu}$ within the muon flavor are weaker than the bounds within the electron flavor from meson decays and LHC data~\cite{Naviliat-Cuncic:2013ylu,Cirigliano:2012ab,Gonzalez-Alonso:2016etj}. Bounds within the muon flavor can also be obtained from the analysis of $pp \to \mu X$ LHC data with missing transverse energy~\cite{Allwicher:2022mcg}. Scalar interactions can be constrained within the assumption of SU$\left(2\right)_L$ gauge invariance of the Standard Model Effective Field Theory from bounds on the electric dipole moments (EDMs)~\cite{Allwicher:2022mcg,ACME:2013pal,Alioli:2018ljm,Griffith:2009zz}.
\end{itemize}

\section{Acknowledgments}

O.T. acknowledges useful conversations with Tanmoy Bhattacharya and Rajan Gupta regarding lattice-QCD results and MINERvA data, with Kevin McFarland regarding the MINERvA data, and with Emanuele Mereghetti regarding up-to-date SMEFT constraints. This work is supported by the US Department of Energy through the Los Alamos National Laboratory and by LANL’s Laboratory Directed Research and Development (LDRD/PRD) program under projects 20210968PRD4, 20210190ER, and 20240127ER. Los Alamos National Laboratory is operated by Triad National Security, LLC, for the National Nuclear Security Administration of U.S. Department of Energy (Contract No. 89233218CNA000001). This work was supported by the U.S. Department of Energy, Office of Science, Office of High Energy Physics, under Award DE-SC0019095. R.J.H. acknowledges support from a Fermilab Intensity Frontier Fellowship. K.B. acknowledges support from the Visiting Scholars Award Program of the Universities Research Association and the Fermilab Neutrino Physics Center Fellowship Program. Fermilab is operated by Fermi Research Alliance, LLC under Contract No. DE-AC02-07CH11359 with the United States Department of Energy. FeynCalc~\cite{Mertig:1990an,Shtabovenko:2016sxi}, Mathematica~\cite{Mathematica}, and MINUIT~\cite{James:1975dr} were extremely useful in this work.

\bibliography{secondclasscurrents}

\end{document}